\documentclass[11pt,a4paper]{article}
\usepackage{latexsym}
\usepackage{amssymb}
\usepackage{amsmath}
\usepackage{amsfonts}
\usepackage{mathrsfs}
\usepackage{cite}
\usepackage{axodraw}
\usepackage{graphics}
\usepackage{graphicx,array}
\usepackage{mathrsfs}
\usepackage{cite}

\catcode`@=11
\renewcommand{\section}{\@startsection{section}{1}{0pt}{\medskipamount}
    {\medskipamount}{\large\bf}} \numberwithin{equation}{section}
\catcode`@=12

\newcommand{\be}{\begin{equation}}
    \newcommand{\ee}{\end{equation}}

\def\tr{{\rm tr}}
\def\tr{{\rm tr}\,}

\def\cN{{\cal N}}

\def\bea{\begin{eqnarray}}
    \def\eea{\end{eqnarray}}
\def\nn{\nonumber}

\def\cN{{\cal N}}

\def\f{\frac}

\def\tr{{\rm tr}\,}

\def\nn{\nonumber}

\def\d{\delta}

\def\g{\gamma}

\def\ve{\varepsilon}

\def\sB{\stackrel{\frown}{\square}}

\def\eq{\eqref}
\def\pr{\partial}
\def\nb{\nabla}

\numberwithin{equation}{section}

\begin{document}
    \begin{titlepage}

        \begin{center}
            \vspace{1cm}

            {\bf \Large On the two-loop divergences in 6D, ${\cal N}=(1,1)$ SYM
                theory}

            \vspace{1.5cm}

            {\bf
                I.L. Buchbinder\footnote{joseph@tspu.edu.ru }$^{\,a,b,c}$,
                E.A. Ivanov\footnote{eivanov@theor.jinr.ru}$^{\,c,d}$,
                B.S. Merzlikin\footnote{merzlikin@tspu.edu.ru}$^{\,a,c,e}$,
                K.V. Stepanyantz\footnote{stepan@m9com.ru}$^{\,f,c}$
            }
            \vspace{0.4cm}

            {\it
                $^a$ Center of Theoretical Physics, Tomsk State Pedagogical University, 634061, Tomsk,  Russia \\ \vskip 0.15cm
                $^b$ National Research Tomsk State University, 634050, Tomsk, Russia \\ \vskip 0.1cm
                $^c$ Bogoliubov Laboratory of Theoretical Physics, JINR, 141980 Dubna, Moscow region, Russia \\ \vskip 0.1cm
                $^d$ Moscow Institute of Physics and Technology, 141700 Dolgoprudny, Moscow region, Russia \\ \vskip 0.1cm
                $^e$ Tomsk State University of Control Systems and Radioelectronics, 634050 Tomsk, Russia\\ \vskip 0.1cm
                $^f$ Department of Theoretical Physics, Moscow State University, 119991 Moscow, Russia
            }
        \end{center}

        \vspace{0.4cm}

\begin{abstract}
            We continue studying $6D, \cN=(1,1)$ supersymmetric Yang-Mills (SYM)
            theory in the $\cN=(1,0)$ harmonic superspace formulation. Using the
            superfield background field method we explore  the two-loop
            divergencies of the effective action in the gauge multiplet sector.
            It is explicitly demonstrated that among four two-loop
            background-field dependent supergraphs contributing to the effective
            action, only one diverges off shell. It is also shown that the
            divergences are proportional to the superfield classical equations
            of motion and hence vanish on shell. Besides, we have
            analyzed a possible structure of the two-loop
            divergences on general gauge and hypermultiplet
            background.
\end{abstract}

    \end{titlepage}

    \section{Introduction}

Supersymmetric field theories in diverse dimensions, especially
those exhibiting the maximally extended supersymmetry, display very
interesting quantum properties. For example, divergences in such
theories sometimes unexpectedly vanish. In some cases such miracles
are caused by a hidden supersymmetry of the theory. This refers,
e.g., to $4D,\,\cN=4$ SYM theory where all possible divergent
diagrams cancel each other due to the maximally extended rigid
$\cN=4$ supersymmetry
\cite{Grisaru:1982zh,Howe:1983sr,Mandelstam:1982cb,Brink:1982pd}.
A consistent derivation of the $4D,\,\cN=2$ non-renormalization theorem
was given in \cite{Buchbinder:1997ib} in $\cN=2$
harmonic superspace formulation
\cite{Galperin:1984av,Galperin:1985ec,Galperin:2001uw}, which is the
most adequate approach to $4D\,\cN=2$ supersymmetric gauge theories.
Another very interesting example of the miraculous divergence
cancelation is provided by $\cN=8$ supergravity, which is the
maximally extended supergravity theory in four dimensions. At
present, it is believed that this theory is finite up to at least
seven loops, see \cite{Bern:2018jmv} and references therein,
although the possible all-loop ultraviolet finiteness is also
discussed (see, e.g., \cite{Kallosh:2010kk,Kallosh:2011dp}).

Similarly to $4D$ (super)gravity theories, the degree of divergence
in higher dimensional gauge theories increases with a number of
loops. One can expect that supersymmetry and, especially, the
maximally extended supersymmetry, is capable to improve the
ultraviolet behavior in such theories. This is the basic reason of
interest in investigating UV divergences of the higher dimensional
supersymmetric gauge theories. They were actually studied for a long
time, see, e.g., \cite{Bern2005,Bern2010,Bern2012,
Fradkin:1982kf,Marcus:1983bd,Marcus:1984ei,Kazakov:2002jd,Howe:1983jm,Howe:2002ui,Bossard:2009sy,
Bossard:2009mn,Bossard:2015dva}. In this paper we will concentrate
on the $6D$ rigid $\cN=(1,1)$ SYM theory. This theory is in many
aspects similar to $4D, \cN=4$ SYM theory in four dimensions, and
one can expect some similarity of the structure of divergences in
both theories. However, they essentially differ in the UV domain. In
contrast to $\cN=4$ SYM theory, which is finite to all loops, its
$6D$ counterpart is non-renormalizable by power-counting.
Nevertheless, the extended supersymmetry leads to the finiteness of
the theory up to two loops, at least on mass shell
\cite{Bork:2015zaa,Kazakov:2002jd,Bossard:2009mn,Bossard:2009sy}.
The modern methods of computing scattering amplitudes
\cite{Bork:2015zaa} demonstrate that UV divergences in
$6D\,\cN=(1,1)$ SYM theory should start from the three-loop level
(see also \cite{Bern2005,Bern2010,Bern2012}).

In our previous works
\cite{Buchbinder:2016url,Buchbinder:2017ozh,Buchbinder:2017gbs,Buchbinder:2017xjb,
Buchbinder:2018bhs,Buchbinder:2019gfb} we studied UV properties of
$6D,\, \cN=(1,0)$ and $\cN=(1,1)$ theories in the $6D$ harmonic
superspace formulation. In particular, it was found that
$6D,\,\cN=(1,1)$ theory is off-shell finite in the one-loop
approximation in the Feynman gauge, although the divergences are
still present in the non-minimal gauges \cite{Buchbinder:2019gfb}
(they vanish on shell). The two-loop divergences in the
hypermultiplet two-point Green functions were shown to also vanish
off shell \cite{Buchbinder:2017gbs}. However, the complete two-loop
calculation in the harmonic superspace approach has not been done so
far. In the present paper we continue the study of $6D,\,\cN=(1,1)$
SYM theory at two loops. We argue that it is not finite off shell in
the Feynman gauge in the two-loop approximation, although the
divergences vanish on shell. Our consideration is limited to the
gauge superfield sector and does not involve the background
hypermultiplet. However, the result is still applicable to other
sectors of the models due to the implicit $\cN=(0,1)$ supersymmetry.
Indeed, we formulate the model in terms of interacting $\cN=(1,0)$
harmonic gauge multiplet and hypermultiplet in adjoint
representation of the gauge group. The action of the model is
manifestly invariant under $\cN=(1,0)$ supersymmetry by
construction. An additional $\cN=(0,1)$ supersymmetry is implicit
and is present only if the hypermultiplet belongs to the adjoint
representation of the gauge group. Note that, albeit $\cN=(1,0)$
theories are in general plagued by anomalies
\cite{Townsend:1983ana,Smilga:2006ax,Kuzenko:2015xiz,Kuzenko:2017xgh},
$\cN=(1,1)$ SYM theory is not anomalous.

The letter is organized as follows. In section
\ref{Section_Harmonic_Superspace} we recall $6D,\,\cN=(1,1)$ SYM
theory in $\cN=(1,0)$ harmonic superspace. Section
\ref{Section_Effective_Action} is devoted to a brief account of the
effective action in the gauge multiplet sector. The effective action
is formulated within the background harmonic superfield method. This
allows us  to perform the calculations in a manifestly gauge
invariant and $\cN=(1,0)$ supersymmetric manner. In section
\ref{Section_Two_Loop} we analyze the structure of possible two-loop
contributions to the effective action and calculate all divergent
terms in this approximation. In section \ref{Section_Hypermultiplet}
we discuss a possible structure of the two loop divergences when the
background hypermultiplet is taken into account. In the last section
\ref{Section_Summary} we summarize the results.

\section{$6D, \cN=(1,1)$ SYM in $\cN=(1,0)$ harmonic superspace}
\label{Section_Harmonic_Superspace}

The six-dimensional maximally extended $\cN=(1,1)$ supersymmetric
gauge theory can be formulated in $\cN=(1,0)$ harmonic superspace.
In this framework it amounts to $\cN=(1,0)$ supersymmetric gauge
theory coupled to the hypermultiplet $q^{+}$ in the adjoint
representation of gauge group. All the necessary notations and
conventions are collected in our previous papers, see, e.g.,
\cite{Buchbinder:2016url,Buchbinder:2017ozh}. Here we  recall only
the basic concepts.

The coordinates of $6D, \cN=(1,0)$ harmonic superspace are denoted
as $(z, u) = (x^M, \theta^a_i,u^{\pm i})$, where $x^M$, $M= 0,..,5$,
are the $6D$ Minkowski space-time coordinates, $\theta^a_i$,
$a=1,..,4\,$, $i=1,2\,$, are Grassmann variables, and $u^{\pm}_i$,
$u^{+i}u^-_i =1\,,$ are the harmonic variables
\cite{Howe:1985ar,Zupnik:1986da}. For the analytic coordinates we
use the notation $\zeta = (x_{\cal A}^M, \theta^{\pm a})$, where
 \be
x^M_{\cal A} \equiv x^M + \frac i2
\theta^{+a}(\gamma^M)_{ab}\theta^{-b},\qquad  \theta^{\pm a}=
u^\pm_k\theta^{ak},
 \ee
and the antisymmetric $6D$ Weyl $\gamma$-matrices are used, $
(\gamma^M)_{ab} = - (\gamma^M)_{ba}\,, (\widetilde{\gamma}^M)^{ab} =
\frac12\varepsilon^{abcd}(\gamma^M)_{cd}\,,$ with  the totally
antisymmetric tensor $\varepsilon^{abcd}$. By definition, the
analytic superfields are annihilated by the spinor covariant
derivative $D_a^+ = u^+_i D^i_a$ and in the analytic basis (where
$D_a^+$ are ``short'') are defined on the analytic harmonic
superspace $(\zeta, u^{\pm i})$.  Also we will need the covariant
derivative $D_a^- = u^-_i D^i_a$ and the harmonic derivatives
 \bea
D^{\pm\pm}= u^{\pm i} \frac{\partial}{\partial u^{\mp i}},\qquad
\qquad D^0 = u^{+i} \frac {\partial}{\partial u^{+i}} - u^{-i} \frac
{\partial}{ \partial u^{-i}}
 \eea
The spinor and harmonic derivatives satisfy the algebra
 \bea
    \{D^+_a,D^-_b\}=i(\gamma^M)_{ab}\partial_M\,, \qquad [D^{++}, D^{--}] = D^0,
    \qquad [D^{\pm\pm},D^{\pm}_a]=0\,, \qquad
    [D^{\pm\pm},D^{\mp}_a]=D^\pm_a\,.
 \eea
The full harmonic and the analytic superspace integration measures
are defined as follows
 \be
d^{14}z \equiv d^6x_{\cal A}\,(D^-)^4(D^+)^4,\quad d\zeta^{(-4)}
\equiv d^6x_{\cal A}\,du\,(D^-)^4 ,
 \ee
where
 \be
(D^{\pm})^4 = -\frac{1}{24} \varepsilon^{abcd} D^\pm_a D^\pm_b
D^\pm_c D^\pm_d.
 \ee

In the harmonic superspace formalism the gauge field is a component
of the analytic gauge superfield $V^{++}$. A necessary ingredient is
also a non-analytic harmonic connection $V^{--}$ obtained as a
solution of the harmonic zero-curvature condition
\cite{Galperin:2001uw}
 \bea
D^{++} V^{--} - D^{--}V^{++} + i[V^{++},V^{--}]=0\,.  \label{zeroc}
 \eea
Using these superfields one can construct the gauge covariant
harmonic derivative $\nb^{\pm\pm}= D^{\pm\pm} + i V^{\pm\pm}$. The
superfield $V^{--}$ is also used to define the spinor and vector
connections in the gauge-covariant derivatives. In the
$\lambda$-frame we have \cite{Ivanov:2005qf}
 \bea
\nb^+_a = D^+_a, \qquad \nb^-_a = D^-_a + i {\cal A}^-_a, \qquad
\nb_{ab} = \pr_{ab} + i {\cal A}_{ab}\,, \label{deriv}
 \eea
where $\nb_{ab} = \f12(\g^M)_{ab} \nb_M$ and
$\nb_M=\partial_M-iA_M\,$, with the superfield connections defined
as
 \be
{\cal A}^-_a= i D^+_a V^{--}\,,\, {\cal A}_{ab} = \f12  D^+_a D^+_b
V^{--}.
 \ee
The covariant derivatives \eq{deriv} satisfy the algebra
\begin{equation}
\{\nb_a^+,\nb^-_b\}=2i\nb_{ab}\,,\qquad
[\nb_c^\pm,\nb_{ab}]=\frac{i}2\ve_{abcd}W^{\pm\, d},\qquad [\nb_M,
\nb_N] = i F_{MN}\,.\label{alg2}
\end{equation}
The superfield  $W^{a\,\pm}$ is the superfield strength of the gauge
multiplet,
 \bea
W^{+a}= -\frac{i}{6}\varepsilon^{abcd}D^+_b D^+_c D^+_d V^{--}, \,\,
W^{-a} = \nb^{--}W^{+a}\,.
 \eea
Also we define the analytic superfield \cite{Bossard:2015dva} $
F^{++} \equiv (D^+)^4 V^{--}$ which satisfies the harmonic
constraint $\nb^{++} F^{++}=0$ following from \eq{zeroc} and the
analyticity of $V^{++}$.

The classical action of $6D,\,\cN=(1,1)$ SYM theory in the harmonic
superspace formulation is written as
    \be
\label{S0} S_0= \tfrac{1}{f^2}\sum\limits^{\infty}_{n=2}
\tfrac{(-i)^{n}}{n} \tr \int d^{14}z\, du_1\ldots du_n
\frac{V^{++}(z,u_1 ) \ldots V^{++}(z,u_n ) }{(u^+_1 u^+_2)\ldots
(u^+_n u^+_1 )} - \tfrac{1}{2 f^2}\tr \int d\zeta^{(-4)}\,
{q}^{+A} \nabla^{++} q^{+}_A\,,
    \ee
where $(u^+_n u^+_1 )^{-1}, \ldots$ are the harmonic distributions
defined in \cite{Galperin:2001uw}, $A=1,2$ is a Pauli-G\"ursey group $SU(2)$ index and $q^{+}_A =\big(q^+, - \tilde{q}^{+}\big), \widetilde{q_A} = \varepsilon^{AB} q^+_B$.
The superfields $V^{++}$ and
$q^{+}_A$ take values in the adjoint representation of the gauge
group, {\it i.e.}, $V^{++}=V^{++\,I}t^{I}\,, q^{+}_A=q^{+\,I}_A t^{I},$
where $t^{I}$ are the gauge algebra generators subjected to the
normalization condition $\tr (t^{I}t^{J}) = \d^{IJ}/2$.  The action
involves the negative-dimension coupling constant $f$, $[f]=m^{-1}$,
and the covariant harmonic derivative
    \be
\nb^{++}q^{+}_A = D^{++}q^{+}_A + i [V^{++},q^{+}_A].
    \ee
The classical equations of motion in the theory have a form
    \bea
F^{++} +\tfrac{i}{2} [ q^{+A}, q^+_A]=0\,, \qquad \nb^{++}q^+_A =0\,.
\label{eqm}
    \eea

The action (\ref{S0}) is invariant under the manifest $\cN=(1,0)$
supersymmetry and an additional hidden $\cN=(0,1)$ supersymmetry.
The hidden supersymmetry mixes the gauge and hypermultiplet
superfields with each other \cite{Bossard:2015dva},
    \be
\delta_{(0,1)} V^{++} = \epsilon^{+ A}q^+_A\,, \quad \delta_{(0,1)}
q^{+}_A = -(D^+)^4 (\epsilon^-_A V^{--})\,, \quad \epsilon^{\pm}_A =
\epsilon_{aA}\theta^{\pm a}\,.\label{Hidden}
    \ee
As a result, the action (\ref{S0}) is invariant under $6D,\, \cN=(1,1)$
supersymmetry. Certainly, it is also invariant under the superfield
gauge transformation
    \be
\d V^{++} = - \nb^{++}\lambda\,, \qquad \d q^{+}_A =
i[\lambda,q^{+}_A]\,, \label{gtr}
    \ee
parameterized by a real analytic superfield $\lambda$.

\section{Effective action}
\label{Section_Effective_Action}

When quantizing gauge theories, it is convenient to use the
background field method allowing to construct the manifestly gauge
invariant effective action. For $6D,\, \cN=(1,0)$  SYM theory in the
harmonic superspace formulation this method was worked out in
\cite{Buchbinder:2016gmc,Buchbinder:2016url,Buchbinder:2017ozh}. In
many aspects it is similar to that for $4D\, \,\cN=2$ supersymmetric
gauge theories \cite{Buchbinder:1997ya,Buchbinder:2001wy} (see also
the review \cite{Buchbinder:2016wng}).

Following the background field method we split the superfield
$V^{++}$ into the sum of the ``background'' superfield $V^{++}$ and
the ``quantum'' one $v^{++}$,
    \be
V^{++}\to V^{++} + f v^{++}.
    \ee
Then we expand the effective action in a power series in quantum
superfields and obtain a theory of the superfields $v^{++},\ q^{+}$
in the background of the classical superfield $V^{++}$, which is
treated as a functional argument of the effective action. Our aim is
to study the two-loop contributions to the effective action in the
gauge superfield sector. To this end, it is sufficient to assume
that the hypermultiplet is purely quantum.

Using the results of refs.
\cite{Buchbinder:2016gmc,Buchbinder:2016url,Buchbinder:2017ozh} the
general expression for the effective action ca be written in the
form
    \bea\label{Effective_Action}
e^{i \Gamma[V^{++}]} = \mbox{Det}^{1/2}\sB \int {\cal
D}v^{++}\,{\cal D}q^+\, {\cal D} b\,{\cal D} c\,{\cal D}\varphi\,
\exp\Big(iS_{\rm total} - \tr\int d\zeta^{(-4)}\, du\, \frac{\delta
\Gamma[V^{++}]}{\delta V^{++}}\, v^{++}\Big),\label{path2}
    \eea
where the operator $\sB=\frac{1}{2}(D^+)^4(\nb^{--})^2$ acting on a
space of analytic superfields is reduced to the covariant superfield
d'Alembertian
    \begin{eqnarray}
\label{Box_First_Part} \sB = \eta^{MN} \nabla_M \nabla_N + i W^{+a}
\nabla^{-}_a + i F^{++} \nabla^{--} - \frac{i}{2}(\nabla^{--}
F^{++}),
    \end{eqnarray}
and $\eta_{MN}$ is $6D$ Minkowski metric with the mostly negative
signature. The total action, $S_{\rm total} = S_{0} + S_{\rm gf} +
S_{\rm  FP} + S_{\rm  NK}\,,$ includes the gauge-fixing term
corresponding to the Feynman gauge,
    \be
S_{\rm gf}[v^{++}, V^{++}] = -\frac{1}{2}\tr \int d^{14}z du_1
du_2\,\frac{v_\tau^{++}(1)v_\tau^{++}(2)}{(u^+_1u^+_2)^2} +
\frac{1}{4}\tr \int d^{14}z du\, v_\tau^{++} (D^{--})^2
v_\tau^{++}\,, \label{SGF}
    \ee
the action for the fermionic Faddeev-Popov ghosts $b$ and $c$, as
well as the action for the bosonic real analytic Nielsen-Kallosh
ghost $\varphi$,
    \bea
S_{FP} &=&-\tr\int d\zeta^{(-4)}\, \nb^{++} b\,
(\nb^{++} c +i[v^{++}, c]), \label{FP}\\
S_{\mbox{\scriptsize NK}} &=& -\frac{1}{2}\tr \int d\zeta^{(-4)}\,
\varphi ({\nb}^{++})^2\varphi \label{NK}.
    \eea
The action \eq{SGF} depends on the background field $V^{++}$ through
the background gauge bridge superfield, in a close analogy  with
$4D$, ${\cal N}=2$ SYM theory.

The calculation of the effective action is carried out in the
framework of the loop expansion. In the one-loop approximation the
quantum corrections to the classical action are determined by the
quadratic part of the action $S_{\rm total}$. After integration over
quantum superfields this quadratic part produces the one-loop
contribution $\Gamma^{(1)}$ to the effective action. The
contributions coming from the Faddeev-Popov ghosts, the
Nielsen-Kallosh ghost, and the quantum hypermultiplet contain
divergences. However, for $\cN=(1,1)$ theory they cancel each other
since in this case the hypermultiplet lies in the adjoint
representation of the gauge group, see refs.
\cite{Buchbinder:2016gmc,Buchbinder:2016url,Buchbinder:2017ozh} for
details. This implies that the theory under consideration is
off-shell finite in the one-loop approximation.

In this paper we will investigate the two-loop divergences. Before
starting the calculations it is instructive to discuss the structure
of propagators and vertices. That part of the total action $S_{\rm
total}$ which is quadratic in quantum superfields defines the
(background-superfield dependent) propagators of these superfields,
which are similar to those for $4D,\ \cN=2$ theory
\cite{Buchbinder:1997ya,Kuzenko:2004gn,Galperin:2001uw}
    \bea
\label{Green2} G^{(2,2)}(\zeta_1,u_1|\zeta_2,u_2)&=&
i<v^{++}(\zeta_1,u_1)
v^{++}(\zeta_2,u_2)> = -2\f{(D^+_1)^4}{\sB}\d^{14}(z_1-z_2)\d^{(-2,2)}(u_1,u_2)\,,\quad\\
\label{Green1} G^{(1,1)}(\zeta_1,u_1|\zeta_2,u_2)&=&
i<q^+(\zeta_1,u_1) \tilde q^+(\zeta_2,u_2)> =
2\f{(D^+_1)^4(D^+_2)^4}{\sB}
\f{\d^{14}(z_1-z_2)}{(u^+_1u^+_2)^3}\,, \\
\label{Green0} G^{(0,0)}(\zeta_1,u_1|\zeta_2,u_2) &=&  i<
b(\zeta_1,u_1) c(\zeta_2,u_2)> =
-(u_1^-u_2^-)G^{(1,1)}(\zeta_1,u_1|\zeta_2,u_2)\,. \vphantom{\Bigg(}
 \eea
In comparison with the $4D, \cN=2$ case, the operator ${\sB}$
has a different form and is given by (\ref{Box_First_Part}).

For calculating the two-loop quantum corrections we will need
vertices which are cubic and quartic in quantum superfields. In the
theory under consideration there are several types of such vertices.

The first type includes the cubic and quartic self-interactions of
the gauge superfield described by the corresponding terms in the
classical action \eq{S0},
    \bea
S^{(3)}_{\rm SYM} &=& \f{i f}{3}\tr\int d^{14} z \prod_{a=1}^{3}
du_a  \f{ v^{++}_1 v^{++}_2 v^{++}_3}{(u^+_1u^+_2)
(u^+_2u^+_3)(u^+_3u^+_1)}\,,
\label{vert3} \\
S^{(4)}_{\rm SYM} &=& \f{f^2}{4}\tr \int d^{14} z \prod_{a=1}^{4}
du_a \f{ v^{++}_1 v^{++}_2 v^{++}_3 v^{++}_4}{(u^+_1u^+_2)
(u^+_2u^+_3)(u^+_3u^+_4)(u^+_4u^+_1)}\,. \label{vert4}
    \eea

The interaction of the gauge multiplet with hypermultiplet can be
also found from classical action \eq{S0} and is given by the term
    \bea
S^{(3)}_{\rm hyper} &=& \f{f}{2}\int d\zeta^{(-4)}\,
f^{IJK}\tilde{q}^+_I v^{++}_J q^+_K\,. \label{vert1}
    \eea

The action \eq{FP} describes the interaction of gauge multiplet and
the Faddeev-Popov ghosts
    \bea
S^{(3)}_{\rm ghost} &=& \f{f}{2}\int d\zeta^{(-4)}\, f^{IJK}
(\nb^{++} b)_I v^{++}_J c_K\,, \label{vert2}
    \eea
where $f^{IJK}$ are the structure constants of the gauge group.

\section{Off-shell two-loop divergences}
    \label{Section_Two_Loop}

Using the power counting \cite{Buchbinder:2016gmc} one can show that
the only possible two-loop divergent contribution in the gauge
superfield sector has the structure
    \bea\label{C_Definition}
\Gamma^{(2)}_{\rm div}[V^{++}] = a \int d\zeta^{(-4)}\, \tr \big(
F^{++} \sB F^{++}\big)\,,
    \label{cont1}
    \eea
where $a$ is a constant, which diverges after removing a
regularization. Below we will calculate the constant $a$ in the
modified minimal subtraction scheme for the considered $\cN=(1,1)$
SYM theory off shell.

In the process of calculation we do not assume any restriction on
the background gauge multiplet and perform the analysis in a
manifestly gauge invariant form. In the two-loop approximation there
are Feynman supergraphs of two different topologies, which we will
call '$\Theta$' and '$\infty$' topologies. The graphs of the
'$\Theta$' topology are generated by cubic interactions. In the
$\cN=(1,1)$ theory under consideration they are presented by eqs.
\eq{vert3}, \eq{vert1}, and \eq{vert2}. The graphs of '$\infty$'
topology contain a vertex corresponding to the interaction. It is
given by eq. \eq{vert4}.
    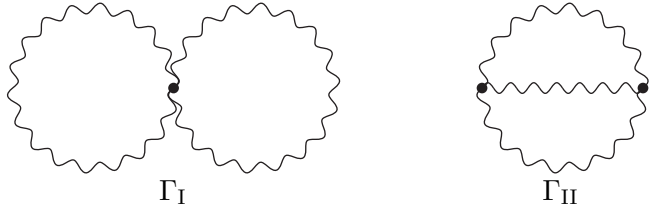
\begin{figure}[tb]
        \begin{center}
            \begin{picture}(220,60)(0,0)
                %%%%%%%%%%%%%%%%%%%%%%%%%%%%%%%%%%%%%%%%%%%%%%%%%%%%
                \PhotonArc(15,40)(30,0,360){2}{18}
                \PhotonArc(75,40)(30,0,360){2}{18} \Vertex(45,40){2}
                \Text(45,1)[]{$\Gamma_{\rm I}$}
                %%%%%%%%%%%%%%%%%%%%%%%%%%%%%%%%%%%%%%%%%%%%%%%%%%%%%%
                \PhotonArc(190,40)(30,0,180){2}{9} \Photon(220,40)(160,40){2}{7}
                \PhotonArc(190,40)(30,180,360){2}{9} \Vertex(160,40){2}
                \Vertex(220,40){2}\Text(190,1)[]{$\Gamma_{\rm II}$}
                %%%%%%%%%%%%%%%%%%%%%%%%%%%%%%%%%%%%%%%%%%%%%%%%%%%
            \end{picture}
            \caption{Two-loop Feynman supergraphs with gauge self-interactions vertices.\label{Fig1}}
        \end{center}
    \end{figure}

\begin{figure}[tb]
        \begin{center}
            \begin{picture}(220,60)(0,0)
                %%%%%%%%%%%%%%%%%%%%%%%%%%%%%%%%%%%%%%%%%%%%%%%%%%%%
                \CArc(45,40)(30,0,180) \Photon(15,40)(75,40){2}{7}
                \CArc(45,40)(30,180,360) \Vertex(75,40){2} \Vertex(15,40){2}
                \Text(45,1)[]{$\Gamma_{\rm III}$}
                %%%%%%%%%%%%%%%%%%%%%%%%%%%%%%%%%%%%%%%%%%%%%%%%%%%%%%
                \DashCArc(190,40)(30,0,180){5} \Photon(220,40)(160,40){2}{7}
                \DashCArc(190,40)(30,180,360){5} \Vertex(160,40){2}
                \Vertex(220,40){2}\Text(190,1)[]{$\Gamma_{\rm IV}$}
                %%%%%%%%%%%%%%%%%%%%%%%%%%%%%%%%%%%%%%%%%%%%%%%%%%%
            \end{picture}
            \caption{Two-loop Feynman supergraphs with hypermultiplet and ghosts
                vertices.\label{Fig2}}
        \end{center}
    \end{figure}
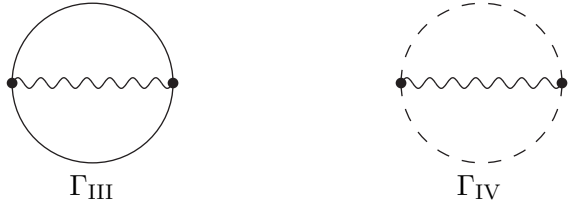

It is convenient to separately consider  the diagrams containing
only the gauge propagators $G^{2,2}$. They are presented in Fig.
\ref{Fig1}, where the gauge propagators are depicted by wavy lines.
Also it is expedient to consider  superdiagrams involving the
hypermultiplet and ghosts propagators together. They are presented
in Fig. \ref{Fig2}. The hypermultiplet propagators $G^{(1,1)}$ are
denoted by solid lines, and the Faddeev--Popov ghost propagators
$G^{(0,0)}$ by dashed lines. In addition, we will take into account
that the theory under consideration is finite at one loop.
Therefore, there is no need to renormalize the one-loop subgraphs in
the two-loop supergraphs\footnote{We emphasize that in the
considered formalism all propagators are background-field
dependent.}.

The analytic expression corresponding to the diagram $\Gamma_{\rm
I}$ (of the `$\infty$' topology) presented in Fig. \ref{Fig1} is
written as
    \bea\label{Infty_Contributon}
&& \Gamma_{\rm I} = - 2 f^2\,{\rm tr} (t^I t^J t^K t^L)
\int {\rm d}^{14}z \int \prod_{a=1}^{4} u_a \nn \\
&& \times \left\{ \frac{G^{(2,2)}_{IJ} (z,u_1 ; z, u_2) \,
G^{(2,2)}_{KL} (z, u_3; z, u_4) } {(u^+_1u^+_2)(u^+_2u^+_3)(u^+_3
u^+_4) (u^+_4 u^+_1)} +\f12 \frac{ G^{(2,2)}_{IJ} (z,u_1 ; z, u_3)
\, G^{(2,2)}_{KL} (z, u_2; z, u_4) } {(u^+_1u^+_2)(u^+_2u^+_3)(u^+_3
u^+_4) (u^+_4 u^+_1)} \right\}. \label{eighth}
    \eea
This expression involves two Green functions $G^{(2,2)}$ in the
coincident $\theta$ limit. According to eq. (\ref{Green2}), each
expression for $G^{(2,2)}$ contains a harmonic $\delta$-function.
Due to these $\delta$-functions the first term in the curly brackets
will contain a singularity, since $(u_1^+ u_2^+)$ and $(u_3^+
u_4^+)$ in the denominator vanish. To avoid this problem, one should
use a 'longer form' of the  gauge superfield Green function
$G^{(2,2)}$ \cite{Galperin:2001uw,Kuzenko:2004gn},
 \be\label{Long_Form}
G^{(2,2)}(\zeta_1,u_1|\zeta_2,u_2) = -(D^+_1)^4\f{1}{(\sB_2)^2}
(D^+_2)^4 (D_2^{--})^2\d^{14}(z_1-z_2)\d^{(-2,2)}(u_1,u_2).
 \ee
Next, in the first term of the expression (\ref{Infty_Contributon})
we should annihilate the Grassmannian delta-functions
$\d^8(\theta_1-\theta_2)|_{\theta_2\to\theta_1}$. This gives the
factor $(u_1^+ u_2^+)^4 (u_3^+ u_4^+)^4$ in the numerator canceling
the singular terms in the denominator. The resulting expression is
proportional to
    \be
\frac{1}{(u_1^+ u_2^+)} \frac{1}{(\sB_2)^2} (u_1^+ u_2^+)^4
(D_2^{--})^2 \d^{(-2,2)}(u_1,u_2)\cdot \frac{1}{(u_3^+ u_4^+)}
\frac{1}{(\sB_4)^2} (u_3^+ u_4^+)^3 (D_4^{--})^2
\d^{(-2,2)}(u_3,u_4).
    \ee

Let us consider a part of this expression depending on $u_1$ and
$u_2$,
    \bea
    && \frac{1}{(u_1^+ u_2^+)} \frac{1}{(\sB_2)^2} (u_1^+ u_2^+)^4 (D_2^{--})^2 \d^{(-2,2)}(u_1,u_2) = \frac{1}{(u_1^+ u_2^+)}
    \Big(\frac{1}{\Box^2} - \frac{1}{\Box^2} i F_2^{++} \nabla^{--} \frac{1}{\Box}\qquad \nn\\
    && - \frac{1}{\Box} i F_2^{++} \nabla^{--} \frac{1}{\Box^2} + \ldots\Big) (u_1^+ u_2^+)^4 (D_2^{--})^2 \d^{(-2,2)}(u_1,u_2),
    \eea
where
    \be
    \Box \equiv \eta^{MN}\nabla_M \nabla_N
    \ee
and $F^{++} = F^{++ A} (T^A_{Adj})$ with $(T^A_{Adj})_{IJ} = - i
f^{AIJ}$. The only possible divergent contributions could appear
from the terms containing $D^{--}$ inside $\nabla^{--}$. However,
    \bea
&&\hspace*{-5mm} \frac{1}{(u_1^+ u_2^+)} D_2^{--} \Big((u_1^+
u_2^+)^4 (D_2^{--})^2 \d^{(-2,2)}(u_1,u_2)\Big)
= D_2^{--}\Big((u_1^+ u_2^+)^3 (D_2^{--})^2 \d^{(-2,2)}(u_1,u_2)\Big)\nn\\
&&\hspace*{-5mm}  + (u_1^+ u_2^-) (u_1^+ u_2^+)^2 (D_2^{--})^2
\d^{(-2,2)}(u_1,u_2) = \Big((D_2^{--})^2 (u_1^+ u_2^+)^2\Big)
\d^{(-1,1)}(u_1,u_2)
    = 2 \d^{(1,-1)}(u_1,u_2).\nn\\
    \eea
Therefore, the first term in eq. (\ref{Infty_Contributon}) diverges.
Taking into account that
    $$
    \mbox{tr}(t^I t^J t^K t^L) f^{AIJ} f^{BKL} = \frac{1}{4}\mbox{tr}([t^I, t^J] [t^K, t^L]) f^{AIJ} f^{BKL} = - \frac{1}{8} (C_2)^2 \delta^{AB},
    $$
we see that (in the Euclidean space after the Wick rotation) its
divergent part is equal to
    \be
4(C_2)^2 \int d^{14}z\, du_1 du_3\, F_{1,\tau}^{++A}
F_{3,\tau}^{++A} \frac{1}{(u_1^+ u_3^+)^2} \cdot \Big[\Big(\int
\frac{d^6k_E}{(2\pi)^6} \frac{1}{k_E^6}\Big)^2\Big]_\infty.
    \ee
The second term in eq. (\ref{Infty_Contributon}) does not contain
harmonic singularities. Therefore, when using the long form of the
gauge propagator, we obtain the expressions proportional to
    \be
(u_1^+ u_3^+)^4 (D_3^{--})^2 \d^{(-2,2)}(u_1,u_3)\cdot (u_2^+
u_4^+)^4 (D_4^{--})^2 \d^{(-2,2)}(u_2,u_4) = 0.
    \ee
Therefore, the contribution of this term vanishes.

The analytic expression for the two-loop diagram $\Gamma_{\rm II}$
(of the `$\Theta$' topology) presented in Fig. \ref{Fig1} is
constructed using the cubic gauge superfield vertex \eq{vert3} and
it has the form
    \bea
\Gamma_{\rm II} &=& -\f{f^2}{6} \int d^{14}z_1
\,d^{14}z_2\prod_{a=1}^{6}
du_a \,f^{I_1J_1K_1}  f^{I_2J_2K_2} \nn \\
&&\times \frac{G^{(2,2)}_{I_1 I_2} (z_1,u_1 ; z_2, u_4) \,
G^{(2,2)}_{J_1 J_2} (z_1, u_2; z_2, u_5) \, G^{(2,2)}_{K_1 K_2}
(z_1,u_3 ; z_2, u_6)} {(u^+_1u^+_2)(u^+_2u^+_3)(u^+_3 u^+_4)
\,(u^+_4u^+_5)(u^+_5u^+_6)(u^+_6 u^+_1)}.
    \eea
As the next steps, we substitute the explicit expression for the
Green function $G^{(2,2)}$ and integrate by parts with respect to
one of the $(D^+)^4$ factors. Also it is possible to calculate the
harmonic integrals over $u_4,u_5,u_6$ using the corresponding
delta-functions which come out from the propagators. As a result, we
obtain
    \bea
\Gamma_{\rm II} &=& -\f{f^2}{6} \int d^{14}z_1
\,d^{14}z_2\prod_{a=1}^{3} du_a \, \f{f^{I_1J_1K_1}
f^{I_2J_2K_2}}{(u^+_1u^+_2)^2(u^+_2u^+_3)^2(u^+_3 u^+_1)^2}
\, (\sB{}^{-1})_{I_1I_2} \delta^{14}(z_1-z_2) \nn \\
&& \qquad\qquad\quad \times (D^{+}_1)^4 \Big( (\sB{}^{-1})_{J_1J_2}
(D_2^+)^4\delta^{14}(z_1-z_2) (\sB{}^{-1})_{K_1K_2}
(D_2^+)^4\delta^{14}(z_1-z_2) \Big).\quad
 \eea
After integrating over $\theta_2$ using the Grassmannian
delta-function we are left with the coincident $\theta_2 \to
\theta_1$ limit in the two remaining delta-functions. In order to
annihilate these Grassmannian delta-functions in the coincident
$\theta$-point limit we need four $(D^\pm)^4$-factors. However we
have only three. The remaining $(D^-)^4$ factor should be obtained
from the expansion of the inverse $\sB$ operator. But in this case
we produce an extra operator $(\partial^2)^4$, so that the overall
momentum degree in the denominator will be $6 + 8 =14$. Taking into
account the presence of the integrations $d^6k d^6q$, we conclude
that the resulting integral is convergent. Therefore, the
superdiagram considered can produce only finite contributions to the
effective action.

Now, let us demonstrate that in $6D,\, \cN=(1,1)$ theory the last
two contributions $\Gamma_{\rm III}$ and $\Gamma_{\rm IV}$ depicted
in Fig. \ref{Fig2} cancel each other. The arguments are basically
analogous to those used for $4D,\, \cN=4$ SYM theory in
\cite{Kuzenko:2004sv}. First, we note that the vertex \eq{vert2}
contains the background-dependent covariant harmonic derivative
$\nb^{++}$, which acts on the ghost field $b$. After integrating by
parts with respect to this derivative, the latter will act on the
ghost propagator $G^{(0,0)}$ which is related to the hypermultiplet
Green function by eq. (\ref{Green0}). Due to this relation the
analytical expression for the sum of two contributions $\Gamma_{\rm
III}$ and $\Gamma_{\rm IV}$ presented in Fig. \ref{Fig2} takes the
form
 \bea
\Gamma_{\rm III}+\Gamma_{\rm IV} &=& f^2\int d\zeta_1^{(-4)}
d\zeta_2^{(-4)}\,
\Big(1+(u^+_1 u^-_2)(u^-_1 u^+_2)\Big)   \nn \\
 && \qquad\qquad\times f^{I_1 J_1 K_1} f^{I_2 J_2 K_2} G^{(2,2)}_{I_1 I_2}(1|2) \, G^{(1,1)}_{J_1 J_2}(1|2) \, G^{(1,1)}_{K_1 K_2}(1|2). \label{Gamma1}
 \eea
    As pointed out in \cite{Kuzenko:2004sv}, the identity $
    1+(u^+_1 u^-_2)(u^-_1 u^+_2) = (u^+_1 u^+_2)(u^-_1 u^-_2)$ allows one
    to transform the contribution \eq{Gamma1} to the form
    \bea
    \Gamma_{\rm III}+\Gamma_{\rm IV}  &=& f^2
    \int d\zeta_1^{(-4)} d\zeta_2^{(-4)}\, (u^+_1 u^+_2)(u^-_1 u^-_2)   \\
    && \qquad\qquad\times f^{I_1 J_1 K_1} f^{I_2 J_2 K_2} G^{(2,2)}_{I_1 I_2}(1|2) \, G^{(1,1)}_{J_1 J_2}(1|2) \, G^{(1,1)}_{K_1 K_2}(1|2) =0\,.
    \eea
This expression vanishes due to the useful property of the harmonic
delta-function $(u^-_1u^-_2) \delta^{(2,-2)}(u_1,u_2) = 0$
\cite{Galperin:2001uw}. Thus, these two diagrams cancel each other.
Obviously, this cancelation takes place only in the case of
$\cN=(1,1)$ theory, when the hypermultiplet is in the adjoint
representation of the gauge group. In a general $6D, \cN=(1,0)$ SYM
theory the diagrams in Fig. \ref{Fig2} enter with different group
factors, which prohibits the cancelation.

Thus, we see that the only divergent contribution comes from the
`$\infty$' superdiagram, and the divergent part of the two-loop
effective action in the gauge superfield sector is given by the
expression
 \be
\Gamma^{(2)}_{\infty, \, \mbox{\scriptsize gauge}} = 8f^2 (C_2)^2
\mbox{tr} \int d^{14}z\, du_1 du_2\, F_{1,\tau}^{++} F_{2,\tau}^{++}
\frac{1}{(u_1^+ u_2^+)^2} \cdot \Big[\Big(\int
\frac{d^6k_E}{(2\pi)^6} \frac{1}{k_E^6}\Big)^2\Big]_\infty.
 \ee
Making use of the identity
    $$
F_{1,\tau}^{++} = \frac{1}{2} D_1^{++} D_1^{--} F_1^{++}\,,
    $$
integrating by parts with respect to the derivative $D_1^{++}$ and
taking into account that $F^{++}_\tau$ is independent of the
harmonic variables, we see that
 \bea
&& \mbox{tr} \int d^{14}z\, du_1 du_2\, F_{1,\tau}^{++}
F_{2,\tau}^{++} \frac{1}{(u_1^+ u_2^+)^2}
= - \frac{1}{2}\mbox{tr} \int d^{14}z\, du_1 du_2\, D_1^{--} F_{1,\tau}^{++} F_{2,\tau}^{++} D_1^{--} \delta^{2,-2}(u_1,u_2)\qquad \nonumber\\
&& =  \frac{1}{2}\mbox{tr} \int d^{14}z\, du\, (D^{--})^2
F_{\tau}^{++} F_{\tau}^{++} = \mbox{tr} \int d\zeta^{(-4)}\, du\,
F^{++} \sB F^{++}.
 \eea

Moreover, in the dimensional regularization scheme we have
 \be
\int \frac{d^D k_E}{(2\pi)^D} \frac{1}{k_E^6} \to
\frac{1}{(4\pi)^{D/2}} \frac{\Gamma(3-D/2)}{\Gamma(3)} =
\frac{1}{(4\pi)^3\varepsilon} +
\frac{1}{128\pi^3}\Big(-\gamma+\ln(4\pi)\Big) + O(\varepsilon).
 \ee
So, in the $\overline{\mbox{MS}}$-scheme,
 \be
\Big[\Big(\int \frac{d^6k_E}{(2\pi)^6}
\frac{1}{k_E^6}\Big)^2\Big]_\infty =
\frac{1}{(4\pi)^6\varepsilon^2}.
 \ee

Thus, the divergent part of the two-loop effective action can be
finally written in the form
 \be
\label{Result} \Gamma^{(2)}_{\infty, \, \mbox{\scriptsize gauge}} =
\frac{8f^2}{(4\pi)^6\varepsilon^2} (C_2)^2 \mbox{tr} \int
d\zeta^{(-4)}\, du\, F^{++} \sB F^{++},
 \ee
and the constant $a$ appearing in Eq. (\ref{C_Definition}) is now
identified as
 \be
a = \frac{8f^2}{(4\pi)^6\varepsilon^2} (C_2)^2. \label{a}
 \ee
An interesting peculiarity of the two loop divergences obtained is
that they contain only leading two-loop pole
$\frac{1}{\varepsilon^2}$, while the sub-leading pole
$\frac{1}{\varepsilon}$ is absent. We believe that the reason for
this may be hidden ${\cal N}=(0,1)$ supersymmetry and the absence of
the off-shell one-loop divergences in the theory under
consideration. The result obtained matches with the statement of
ref. \cite{Bossard:2015dva} that the candidate two-loop counterterms
in $\cN=(1,1)$ SYM theory vanish on mass shell, provided they are
required to be $\cN=(1,0)$ off-shell supersymmetric and gauge
invariant. More details on this point are given in the next section.

    \section{Hypermultiplet dependence of the two-loop divergences}
    \label{Section_Hypermultiplet}

In the previous section we have calculated the two-loop divergences
in the gauge multiplet sector, where the background hypermultiplet
$q^{+}$ is absent. Now we discuss a possible structure of the two-loop divergences in the case when the background hypermultiplet is
taken into account. Of course, the hypermultiplet-dependent
contribution to two-loop divergences can be obtained by the
straightforward quantum computations of the two-loop effective
action. However, the general form of such divergences can in
principle be described without direct calculations, just starting from the expression (\ref{Result}) and assuming the invariance of the effective action under
the hidden $\cN=(0,1)$ supersymmetry. Taking into account the result
(\ref{Result}) one might expect that including the on-shell background
hypermultiplet will merely lead to the replacement of $F^{++}$ in (\ref{Result}) by the
total classical equation of motion for the background gauge multiplet \eqref{eqm} coupled to hypermultiplet.

As was proved in \cite{Bossard:2015dva,Smilga2016}, only the classical
action (\ref{S0}) is $\cN=(1,1)$ supersymmetric off shell, while in any other $\cN=(1,1)$ invariant the hidden $\cN=(0,1)$ supersymmetry must be on-shell.
Therefore we will assume here that the hypermultiplet satisfies the classical
equations of motion
\bea
\nabla^{++} q^{+}_A = \nabla^{--} q^{-}_A = 0\,,
\label{equationq}
\eea
where $q^{-}_A := \nabla^{--}q^{+}_A\,.$ In this case the
$\cN=(0,1)$ supersymmetry transformation \eqref{Hidden} for
non-analytic gauge potential takes the form
\bea
 \delta_{(0,1)} V^{--} = \epsilon^{-A} q^{-}_A, \quad \epsilon^{- A} = \epsilon^{A}_a \theta^{- a}.
\eea

Let us now rewrite the expression \eqref{cont1} in the central basis,
\begin{equation}
\Gamma^{(2)}_{\rm div}[V^{++}] = -\f{a}{2} \int d^{14}z du\, \tr \,
\big(\nabla^{--} F^{++}\big)^2.
 \label{4}
\end{equation}
Here we made use of the definition of covariant d'Alembertian
\eqref{Box_First_Part} and integrated by parts with respect to the harmonic
derivative $\nb^{--}$. The coefficient $a$ is given by  eq. (\ref{a}). Our
aim is to find the appropriate terms which should be added to the
action \eqref{4} to ensure the invariance under hidden $\cN=(0,1)$ supersymmetry
transformations. First, we rewrite the $\cN=(0,1)$ transformation \eq{Hidden} in the form
\begin{eqnarray}
\delta_{(0,1)} F^{++} &=& -i\epsilon^A_b [W^{+ b}, q^+_A] - i[\epsilon^{-A} q^+_A, F^{++}]\,, \nonumber \\
\delta_{(0,1)} q^{+A} &=& \epsilon^A_b W^{+ b} - i[\epsilon^{-B} q^+_B, q^{+A}] - \epsilon^{-A} E^{++}\,,\nonumber \\
\delta_{(0,1)} q^{-A} &=& \epsilon^A_b W^{- b}- i[\epsilon^{-B} q^+_B, q^{-A}] - \epsilon^{-A}\nabla^{--} E^{++}\,,
\label{3}
\end{eqnarray}
where $E^{++} := F^{++} + \tfrac{i}2[q^{+A}, q^+_A]$.  After that one can see that the following generalization
of the action \eq{4},
\be
\Gamma^{(2)}_{\rm div}[V^{++},q^+] = -\f{a}{2}\int d^{14}z du\, \tr \Big\{  \big(\nabla^{--} F^{++}\big)^2 -
2i [q^{-A}, q^-_A] F^{++} + \frac12 [q^{-A}, q^-_A][q^{+ C}, q^+_C] \Big\}, \label{9}
\ee
for the background hypermultiplet satisfying (\ref{equationq}), under \eq{3} is  transformed as
\be
\delta \Gamma^{(2)}_{\rm div}[V^{++},q^+] =-\f{a}{2}\int d^{14}z du\, 4i\,\epsilon^{-A} \tr q^-_A[E^{++}, \nabla^{--} E^{++}]
\ee
and so is invariant modulo the gauge superfield equation of motion $E^{++} = 0$. The  action \eqref{9} can be rewritten, up to a total harmonic
derivative, as
\be
\Gamma^{(2)}_{\rm div}[V^{++},q^+] =-\f{a}{2} \int d^{14}z du\, \tr\big(\nabla^{--} E^{++}\big)^2\,.
\label{10}
\ee
Passing to the analytic basis, we finally obtain
\bea
\Gamma^{(2)}_{\rm div}[V^{++},q^+]= a\int d \zeta^{(-4)} \tr
E^{++}\sB E^{++},
\label{cont4}
\eea
%where the hypermultiplet equations of motion are assumed.
We see that two-loop divergences vanish on
the total  mass-shell \eqref{eqm}, as expected.

Finally, we note that the superficial degree of divergence in
$\cN=(1,0)$ SYM theory was calculated in \cite{Buchbinder:2016gmc} in the form
 \bea
\omega=2L-N_q -\f12 N_D\,, \label{omega}
 \eea
where $L$ is a number of loops in the supergraph with $N_q$ external
lines of hypermultiplet and $N_D$ is a number of spinor derivatives
acting on the external lines. Divergent contributions correspond to
the case $\omega \geqslant 0$. Hence at $L=2$, the number of the
external hypermultiplet lines should be $N_q \leqslant 4$. Possible
divergent contributions in the gauge superfield sector at two loops
have the universal structure \eq{cont1}. The number of external
hypermultiplet lines should be even to secure gauge invariance.
Hence the possible hypermultiplet-dependent divergent contributions
have two or four  external hypermultiplet lines. Taking into account these reasonings
and $\cN=(1,0)$ supersymmetry, we obtain the following
expression for the two-loop divergences
 \bea
\label{cont2} \Gamma^{(2)}_{\infty}[V^{++},q^+]= a\int d
\zeta^{(-4)} \tr \Big(F^{++}\sB F^{++} + i c_1 F^{++} \sB [q^{+A}, q^+_{A}] +c_2[q^{+A}, q^+_A]\sB[q^{+ B}, q^+_B] \Big) \\
    \nonumber
    +\,\, \text{terms proportional to the hypermultiplet equations of motion},
 \eea
where the constant $a$ ia given by (\ref{a}) and $c_1, c_2$ are the
arbitrary dimensionless numerical coefficients, which can be fixed
only within the quantum field theoretical computations of the
effective action. Comparing (\ref{cont2}) with (\ref{9}), we observe
that the role of hidden ${\cal N}=(0,1)$ supersymmetry is just to
relate the unknown constants $c_1$ and $c_2$ to the original
constant $a$. Indeed, requirement of invariance of the expression
(\ref{cont2}) under the ${\cal N}=(0,1)$ supersymmetry yields the
same expression (\ref{cont4}).

\section{Summary}
\label{Section_Summary}

In the present paper we have studied two-loop divergent
contributions to the effective action for $6D, \,\cN=(1,1)$ SYM
theory formulated in  $\cN=(1,0)$ harmonic superspace. In this
approach it amounts to the model \eq{S0} of the minimally coupled
$\cN=(1,0)$ gauge multiplet and the hypermultiplet, both in the
adjoint representation of the gauge group. The classical action of
the model is invariant under an additional $\cN=(0,1)$
supersymmetry, so that it actually describes $\cN=(1,1)$ SYM
theory.

In the papers \cite{Buchbinder:2016url,Buchbinder:2016wng} we
have demonstrated by explicit calculations that, in the minimal gauge,
$\cN=(1,1)$ SYM theory in six-dimensions is one-loop finite {\it
off} shell. In the present paper, using the superfield background
field method, we have calculated the divergent part of the two-loop
effective action in the gauge multiplet sector. The corresponding
background field dependent supergraphs determining the effective
action are given by Fig. \ref{Fig1} and Fig. \ref{Fig2}. It was
shown that the divergences of the supergraphs $\Gamma_{III}$ and
$\Gamma_{IV}$ in Fig. \ref{Fig2} cancel each other due to the hidden
$\cN=(0,1)$ supersymmetry. The supergraph $\Gamma_{II}$ in Fig.
\ref{Fig1} is finite. The total divergence is only due to the supergraph
$\Gamma_{I}$ in Fig. \ref{Fig1}. The corresponding divergent
contribution to the two-loop effective action is proportional to the
classical equation of motion. This means that the theory is not
off-shell finite at two loops in the gauge multiplet sector even in
the Feynman gauge, while the divergences vanish on shell in this
sector. Nevertheless, it is worth pointing out that the two-loop
divergences in the theory under consideration are 'softer' in some
sense as compared with the general quantum field theory setting. The
divergent part of the two-loop effective action (\ref{Result})
contains only the leading two-loop pole $\frac{1}{\varepsilon^2}$,
the sub-leading pole $\frac{1}{\varepsilon}$ being absent. This
peculiarity could be attributed to hidden ${\cal N}=(0,1)$
supersymmetry.

Also, we have analyzed, on the grounds of gauge invariance, power
counting, the explicit $\cN=(1,0)$ supersymmetry and the hidden $\cN=(0,1)$
supersymmetry, the possible structure of the two-loop divergences for
$\cN=(1,1)$ super Yang-Mills theory in an arbitrary gauge and
hypermultiplet background. It was shown that such divergences vanish
on the total equations of motion \eqref{eqm} and contain an arbitrary
dimensionless numerical coefficient. To fix this coefficient, we
must carry out the direct quantum field theoretical calculations.
Thus, obviously, the most urgent  problem for further study is to calculate the two-loop
divergences in the general background
field setting, including not only the background gauge multiplet but
the background hypermultiplet  as well. We hope to confirm our
assertion that the total two-loop divergences involve the complete
classical equation of motion.

Another interesting problem is to calculate the two-loop divergences
for the general $\cN=(1,0)$ SYM theory without hidden ${\cal
N}=(0,1)$ sector.   We plan to perform the detailed calculation of
the two-loop divergent contributions for the general $\cN=(1,0)$
gauge theory in a forthcoming work.

\section*{Acknowledgements}
The work is partially supported by Russian Scientific Foundation,
project No 21-12-00129.

\end{document}